# Numerical Simulations of Magnetized Astrophysical Jets and Comparison with Laboratory Laser Experiments

V. S. Belyaev[1], G. S. Bisnovatyi-Kogan[2,3], A. I. Gromov[4], B. V. Zagreev[1], A. V. Lobanov[1], A. P. Matafonov[1], S. G. Moiseenko[2], and O. D. Toropina[2]

[1]Central Research Institute forMachine Building (TsNIIMash),
Korolev, Moscow region, 141070 Russia
[2]Space Research Institute, Russian Academy of Sciences, Moscow, 117997 Russia
[3]Moscow Engineering Physical Institute, National Nuclear Research Institute,
Moscow, 115409 Russia
[4]Lebedev Physical Institute, Moscow, 119991 Russia

**Abstract**—The results of MHD numerical simulations of the formation and development of magnetized jets are presented. Similarity criteria for comparisons of the results of laboratory laser experiments and numerical simulations of astrophysical jets are discussed. The results of laboratory simulations of jets generated in experiments at the Neodim laser installation are presented.

## 1. INTRODUCTION

Astrophysical jets (directed plasma ejections) are observed on various astrophysical scales and are encountered in astrophysical objects ranging from active galactic nuclei and quasars to compact objects in stellar binary systems and young stellar objects. The appearance of jets during explosions of collapsing supernovae is also possible. It is currently believed that a possible reason for the origin of cosmic gamma-ray bursts is an anisotropic explosion in which narrowly directed jets are formed. The formation and evolution of relativistic jets is one of the most interesting problems in modern astrophysics [1]. Studies of relativistic jets are being conducted in two main directions: using observations in the optical, X-ray, and radio, and using multi-dimensional MHD numerical simulations. Such simulations should ideally take into account the complex nature of relativistic flows, including the gravitation and magnetic field of the central object. Methods in laboratory astrophysics have recently been added to the approaches used to study relativistic jets. These enable the creation of a relativistic stream of plasma using lasers, and enable studies of the morphology, evolution, and structural features of astrophysical jets under laboratory conditions. To enable comparisons of the results of laboratory experiments with astrophysical observations and numerical experiments, methods incorporating principles of similarity must be used. The magnetohydrodynamical (MHD) equations describing the formation and evolution of jets are non-linear, even in the one-dimensional case. A theoretical description of the processes occurring in jets requires application of the methods of multi-dimensional numerical simulations. Multi-dimensional numerical computations can be used to obtain a detailed picture of the matter flows in a jet and the shock–wave structure of the flow. Numerical modeling enables computations for input data corresponding to the parameters of both astrophysical jets and laboratory experiments. Laboratory studies are being undertaken by several research groups. In 2014, Fua et al. [2] began work on the creation of a magnetized, supersonic jet under laboratory conditions. A ring-shaped laser beam was used to heat a flat target, and an ejection of matter was observed. Numerical simulations using the two-dimensional FLASH code were carried out in parallel with the laboratory experiment. Two main configurations were studied, with the laser beams incident on the target having the shape of a simple circular spot and a ring. In the simulations with the first configuration, the density and pressure gradient proved to be insufficient to create a directed ejection of matter. In the simulations with the configuration, the ring-like laser beam was able to give rise to a supersonic ejection of matter with high density, temperature, and velocity, and a high degree of collimation. A strong toroidal magnetic field was observed in the outflowing jet. Experiments on laboratory modeling of astrophysical jets have also been conducted at the PF-3 plasma focus installation of the Kurchatov Institute [3]. Results of mathematical modeling of these experiments are presented in [4]. The results of a laboratory laser experiment and numerical simulations of a magnetized jet flowing out from a young stellar object are presented in [5]. The jet collimation in these computations is achieved due to the presence of a poloidal field. Experimental studies on modeling of astrophysical jets and astrophysical shocks at the Laboratoire pour l'Utilisation des Lasers Intenses (LULI) and other French facilities are considered in [6–12].

## 2. OBSERVATIONS OF RELATIVISTIC DIRECTED EJECTIONS (JETS)

The radio (and in some cases, optical and X-ray) maps of many quasars and active galactic nuclei (AGNs) display bright, compact central features from which one or two jets emerge. In high angular resolution images, these jets are resolved into bright knots separated by relatively dimmer regions. High degrees of linear polarization are observed, sometimes exceeding 50%, which can be understood if we are observing synchrotron radiation by relativistic electrons in weak, but ordered magnetic fields. The estimated lifetimes of these electrons based on the observed luminosity and spectrum often yield values appreciably shorter than the kinematic time scale $d/c$, where $d$ is the distance from the point at which an electron radiates to the central source and $c$ is the speed of light. Since the jet presumably arises as the result of an ejection from the central source, or a continuous outflow from that source, this observation leads to the need for continuous, or in situ, acceleration of the electrons in jets. The best known objects in which jets are visible are the core of the giant elliptical galaxy M87 in the constellation Virgo [13, 14] and the first quasar discovered, 3C 273 [15]. Relativistic motion in jets can explain the frequently observed one-sidedness of the jets. It is believed that the jet emerging in the opposite direction is not visible due to relativistic effects. At the same time, the brightness of the visible jet, which is moving toward the Earth at relativistic speed, is enhanced. The jets in M87 and 3C 273 are both onesided. In principle, the observed one-sided nature of many jets could also reflect a real situation in which the ejections from the core occur asymmetrically. An example of a symmetrical two-sided jet is given by the radio galaxy IC 4296 [16]. Irregular variability of the luminosities and spectral energy distrbutions of quasars and AGNs from the radio to the gamma-ray is observed on various time scales. Observations carried out over several years show that 80–100% of AGNs selected using other methods (spectroscopy, color selection, enhanced UV emission) are variable.

## 3. ANALYSIS OF SIMILARITY CRITERIA

The scales of physical processes occurring in astrophysical jets differ strongly from those in jets created due to the action of a laser pulse on a target in a laboratory. However, the application of similarity criteria make it possible to obtain correspondence between a number of characteristics of astrophysical and laboratory jets. The application of such similarity criteria is useful for direct comparisons of laboratory data and both observations and multi-dimensional numerical simulations. Investigations of similarity criteria for comparisons of the results of laboratory experiments and astrophysical processes have been carried out in many studies (see, e.g., [17–20]). In spite of the very different physical parameters of laboratory and astrophysical jets, there exist scaling criteria that can be used to relate one to another .Following [21], we will consider the following system of one-dimensional equations for the non-stationary hydrodynamics of a compressible fluid [22]:

$$\frac{\partial \rho}{\partial t} + \nabla \cdot (\rho v) = 0, \quad dM = \rho dV, \qquad (1)$$

$$\rho \frac{dv}{dt} = -\nabla P, \quad \frac{d}{dt} = \left[\frac{\partial}{\partial t} + (v \cdot \nabla)\right], \qquad (2)$$

$$\frac{dP}{dt} - \gamma \frac{P}{\rho} \frac{d\rho}{dt} = -(\gamma - 1)\Lambda(\rho, T), \qquad (3)$$

where $t$, $v$, $M$, $V$, $\rho$, $P$, $\gamma$, and $\Lambda(\rho, T)$ are the time, velocity, mass, volume, density, pressure, adiabatic index, and cooling function, respectively. For simplicity, we assumed that the cooling function $\Lambda(\rho, T)$ can be represented in the form

$$\Lambda(\rho, P) = \Lambda_0 \rho^\epsilon p^\varsigma.$$

We wrote the equation of state in the form

$$p(\rho, T) = C_{\text{EOS}} \rho^\mu p^\nu.$$

We wrote the equation for the internal energy in the form

$$\rho e = \frac{P}{\gamma - 1},$$

where $e$ is the specific internal energy. Following [21], we determined appropriate similarity criteria by writing the following relations:

$$x = a^{\delta_x} \hat{x}, \quad t = a^{\delta_t} \hat{t}.$$

As independent variables of the hydrodynamical equations, we have here the new coordinate and time $\hat{x}$ and $\hat{t}$. Similarly, we have for the dependent variables

$$v = a^{\delta_v} \hat{v}, \quad \rho = a^{\delta_\rho} \hat{\rho}, \quad P = a^{\delta_P} \hat{P}.$$

We obtain for the physical parameters

$$C_{\text{EOS}} = a^{\delta_{C_{\text{EOS}}}} \hat{C}_{\text{EOS}}, \quad \Lambda_0 = a^{\delta_{\Lambda_0}} \hat{\Lambda}_0.$$

Writing this in general form,

$$A_i = a^{\delta_i}, \quad i = x, t, v, p, \rho, \Lambda.$$

Substituting these values into the equation of continuity
Yields

$$a^{\delta_\rho - \delta_t} \frac{\partial}{\partial \hat{t}} \hat{\rho}(\hat{x}, \hat{t})$$
$$+ a^{-\delta_x + \delta_\rho + \delta_v} \frac{\partial}{\partial x} \left[ \hat{\rho}(\hat{x}, \hat{t}) \hat{v}(\hat{x}, \hat{t}) \right] = 0.$$

Canceling out the factor of $\delta\rho$ and multiplying by the factor $a\delta t$ yields

$$\frac{\partial}{\partial \hat{t}} \hat{\rho}(\hat{x}, \hat{t})$$
$$+ a^{-\delta_x + \delta_t + \delta_v} \frac{\partial}{\partial x} \left[ \hat{\rho}(\hat{x}, \hat{t}) \hat{v}(\hat{x}, \hat{t}) \right] = 0.$$

In order for the equation of continuity to remain unchanged in the transition to other variables, we must have $\delta_v + \delta_t - \delta_x = 0$. It follows that

$$A_v = A_x / A_t.$$

Applying the same procedure for the equation of motion
Yields

$$\left( \frac{\partial}{\partial \hat{t}} + \hat{v} \frac{\partial}{\partial \hat{x}} \right) \hat{v}(\hat{x}, \hat{t})$$
$$= -a^{\delta_p - \delta_\rho - 2(\delta_x - \delta_t)} \frac{1}{\hat{\rho}} \frac{\partial}{\partial \hat{x}} \hat{P}(\hat{x}, \hat{t}).$$

To ensure invariance of the equation of motion, we must have

$$\delta_p - \delta_\rho = 2(\delta_x - \delta_t) = 2\delta_v,$$

and consequently,

$$A_p = A_\rho (A_v)^2 = A_\rho (A_x / A_t)^2.$$

We obtain similarly for the energy equation

$$\left(\frac{\partial}{\partial \hat{t}} + \hat{v}\frac{\partial}{\partial \hat{x}}\right) \hat{P}(\hat{x}, \hat{t}) - \gamma\frac{\hat{P}}{\hat{\rho}}\left(\frac{\partial}{\partial \hat{t}} + \hat{v}\frac{\partial}{\partial \hat{x}}\right) \hat{\rho}(\hat{x}, \hat{t}) = -a^{\delta_\Lambda + \delta_t - \delta_P}(\gamma - 1)\hat{\Lambda}(\hat{\rho}, \hat{P}),$$

**Table 1.** Comparison of the parameters of laboratory and astrophysical jets

| Parameter | Neodim laboratory jet [23] | Laboratory jet after scaling | Jets from young stars |
|---|---|---|---|
| $x$, cm | 0.1–1 | $(3-30) \times 10^{17}$ | $\sim 3 \times 10^{17}$ |
| $t$, s | $10^{-8} - 10^{-7}$ | $(3-30) \times 10^{9}$ | $\sim 3 \times 10^{10}$ |
| $v$, cm/s | $10^7$ | $10^7$ | $(2-5) \times 10^7$ |
| $\rho$, g/cm$^3$ | $10^{-4}$ | $10^{-23}$ | $10^{-23}$ |
| $n$, cm$^{-3}$ | $2 \times 10^{19}$ | 2 | 1–100 |
| $T$, K | $10^4$ | $10^4$ | $(1-6) \times 10^4$ |
| $H$, G | $10^8 - 10^9$ | 1–10 | 1–100 |

where we have assumed that the adiabatic index is invariant. The energy equation will be invariant when $\delta_\Lambda = \delta_P - \delta_t$, which implies that

$$A_\Lambda = A_P / A_t.$$

In the presence of a magnetic field, the system of MHD equations is supplemented by Maxwell's equations. The equation of motion is written in the form (in the case of ideal MHD)

$$\rho\frac{dv}{dt} = -\nabla P - \frac{1}{4\pi} H \times \nabla \times H.$$

Introducing the scaling relation for the magnetic field

$$H, H = a_H^{\delta_H} \hat{H},$$

we obtain

$$A_H = \sqrt{8\pi A_p}.$$

Thus, we have four relations relating the seven scaling coefficients $A_i$. Consequently, three of these coefficients can be chosen independently.

### 3.1. Application of Similarity Criteria to Test Models for Relativistic Astrophysical Processes and Cosmology using a Laser Installation

We used the similarity relations presented above to compare the parameters of jets produced in laser experiments with those of astrophysical jets. The characteristic values for length, time, and density for jet outflows from young stars are $x = 0.1$ pc $= 3 \times 10^{17}$ cm, $t = 1000$ yrs ($3 \times 10^{10}$ s), and $\rho = 10^{-24}$–$10^{-22}$ g/cm$^3$. Accordingly, we chose as independent scaling coefficients [21]:

$$A_x = A_t = 3 \times 10^{18}, \quad A_\rho = 10^{-19}.$$

**Table 2.** Comparison of the parameters of a Neodim laboratory jet with the jet of an AGN

| Parameter | Laboratory jet after scaling | AGN jet |
|---|---|---|
| $x$, cm | $(0.3–3) \times 10^{18}$ | $3 \times 10^{18}$ |
| $t$, s | $(0.3–3) \times 10^{9}$ | $10^{8}$ |
| $v$, cm/s | $10^{9}$ | $3 \times 10^{10}$ |
| $\rho$, g/cm$^3$ | $10^{-26}$ | $10^{-26}$ |
| $n$, cm$^{-3}$ | $10^{-2}$ | $10^{-2}$ |
| $T$, K | $10^{11}$ | $10^{11}$ |
| $H$, G | $10^{-1}$ | $10^{-3}$ |

Applying the scaling with these coefficients yields the values for the length, time, and density for jets from young stars given in Table 1. This scaling yields a satisfactory agreement between the parameters of jet outflows from young stars and a jet produced at the Neodim facility. We then chose the scaling coefficients

$$A_x = 3 \times 10^{18}, \quad A_t = 3 \times 10^{16}, \quad (4)$$
$$A_\rho = 10^{-22},$$

applied these to the parameters of a laboratory jet produced at the Neodim facility, and compared the resulting values to the parameters of jets from AGNs (see, e.g., [24]). The results of this comparison are given in Table 2; in this case, as well, there is a satisfactory agreement between the parameters of jets ejected from AGNs and the experimental parameters of the jet produced in a laboratory laser experiment.

## 4. NUMERICAL MODELING OF MAGNETIZED JETS

### 4.1. General Formulation of the Problem

The aim of our study is the *numerical* modeling and analysis of the results of laboratory experiments designed to imitate the propagation of astrophysical relativistic streams of plasma (jets). The laboratory experiment consisted of the following. A foil of Cu or Ta with a thickness of 30–50 $\mu$m was placed in a cylindrical chamber and used as a target. An energy of 10 J = 6.24 × $10^{19}$ eV was uniformly and instantaneously applied to a central region of the target with a diameter of 10 $\mu$m. As a result of heating of the target, two directed plasma ejections arose, having the form of symmetrical jets. Only one of these was considered in the numerical simulations. Figure 1 shows a schematic of the experiment, and the experiment is described in detail in Section 5.

We did not consider the processes of laser heating, melting, and vaporization of the target in detail in the simulations. A large number of studies have been dedicated to laser-ablation processes. For example, Mazhukin et al. [25, 26] investigated nanosecond laser ablation in the subcritical and transcritical regimes. They also conducted simulations of laser induced explosive boiling of a metal [27]. We assumed that the ablation had already taken place and the material had vaporized from the surface of the target, and took this into account when determining the initial and boundary conditions.

We used the system of MHD equations [28] in our numerical simulations of the experiment. Obtaining a more exact description of the processes occurring in the target material as a result of irradiation by a powerful laser requires combining an MHD and kinetic approach. However, we can restrict our treatment to the system of MHD equations in the initial stage of the simulations. We considered two cases in the simulations. In the first, we did not include an external magnetic field. In the second, we included a constant external magnetic field having two configurations: a poloidal magnetic field directed along the normal to the target, and a toroidal field. Simulations were also carried out for various magnetic-field strengths. In both cases, the ejection of plasma from the target due to the action of the laser radiation was modeled. We analyzed the parameters of the resulting stream of one-component plasma: the shape of the flow at various distances from the target and at various times, the spatial distribution of the density and energy of the plasma, and the magnetic field, when included.

### 4.2. Simulation Region

Analysis of the experiment showed the formation of a jet due to heating of the foil, with the geometry of the incident laser beam not being important. This means that the mathematical model for the formation of the jet in the experiment shown schematically in Fig. 1 can be constructed in an axially symmetric approximation, assuming that the laser-heated spot is circular. We used inertial cylindrical coordinates ($r$, $\varphi$, $z$) in the modeling, with the coordinate origin located at the middle of the target. The orientation of the coordinate system was chosen so that the $z$ axis was parallel to the poloidal magnetic field $H$ and perpendicular to the target. We assumed axial symmetry in the distributions of all the macroscopic quantities $\rho$, $T$, $v$, and $H$, with $\partial/\partial\varphi = 0$; however, all three components of the velocity $v$ and magnetic field $H$ were computed.

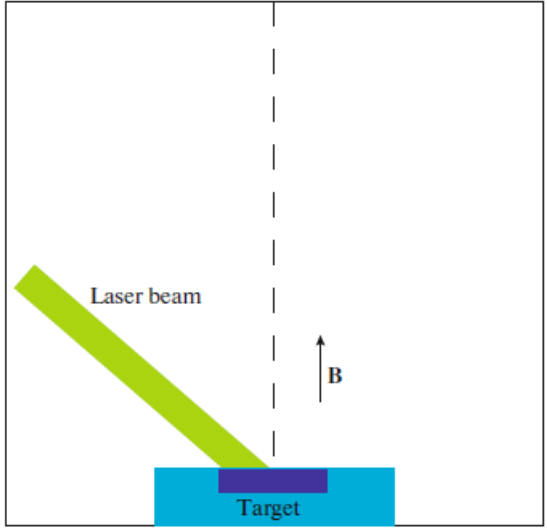

**Fig. 1.** Schematic of the experiment.

The simulations were carried out in half the $r-$ coordinate plane. The computations used a uniform difference grid $(r, z)$ with the number of cells being $N_R \times N_Z = 257 \times 513$ or $N_R \times N_Z = 513 \times 1026$; the coordinates of the grid nodes lay in the ranges $0 \leq R \leq R_{max}$, $0 \leq z \leq Z_{max}$. The dimensions of the simulation region were $Z_{max} = 500$ $\mu$m, $R_{max} = 250$ $\mu$m. The target (central part of the foil, where the laser energy was applied) was treated as a uniform cylinder with radius $R_d$ and thickness $Z_d$, with $R_d \ll R_{max}, Z_{max}$, and $Z_d \ll R_{max}, Z_{max}$. In the computations, we adopted the value $R_d = 5\mu$m, with the radial cell size being $DR = R_{max}/N_R = 0.96$, or $0.48$ $\mu$m. Figure 2 presents a schematic of the simulation region.

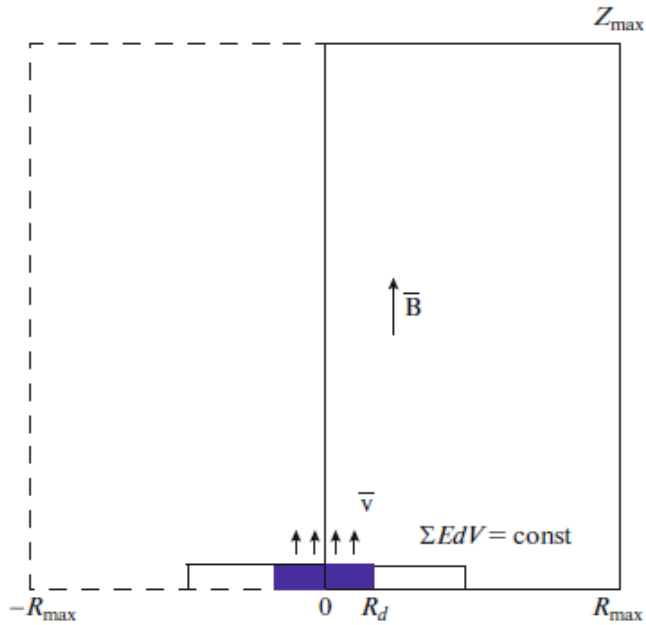

**Fig. 2.** Simulation region.

*4.3. System of Equations*

We described the processes occurring in the target material upon instantaneous heating by the laser beam and simulated the resulting plasma flow using a standard system of MHD equations taking into account the finite conductivity of the target material [28] and the absence of appreciable gravitation:

$$\frac{\partial \rho}{\partial t} + \nabla \cdot (\rho \mathbf{v}) = 0, \quad (5)$$

$$\rho \frac{\partial \mathbf{v}}{\partial t} + \rho(\mathbf{v} \cdot \nabla)\mathbf{v} = -\nabla p + \frac{1}{c}\mathbf{J} \times \mathbf{H}, \quad (6)$$

$$\frac{\partial \mathbf{H}}{\partial t} = \nabla \times (\mathbf{v} \times \mathbf{H}) + \frac{c^2}{4\pi\sigma}\nabla^2 \mathbf{H}, \quad (7)$$

$$\frac{\partial(\rho\varepsilon)}{\partial t} + \nabla \cdot (\rho\varepsilon\mathbf{v}) = -p\nabla \cdot \mathbf{v} + \frac{\mathbf{J}^2}{\sigma}, \quad (8)$$

$$\frac{4\pi}{c}\mathbf{J} = \nabla \times \mathbf{H}. \quad (9)$$

Here, $c$ is the speed of light, $\rho$ the density of the matter, $v$ the velocity, $H$ the magnetic-field strength, $\varepsilon$ the internal energy per unit mass of the matter, $\mathbf{J}$ the electrical current density, and $\sigma$ the effective electrical conductivity of the medium. We used the adiabatic equation of state for an ideal gas to describe the thermodynamical properties of the matter:

$p = (\gamma - 1)\rho\varepsilon,$

where $\gamma$ is the adiabatic index. We used the value $\gamma = 5/3$ in these computations, which corresponds to the standard adiabatic index for a monatomic gas. When writing the system of equations (5)–(9), we took into account Ohm's law in the form

$\mathbf{J} = \sigma(\mathbf{E} + \mathbf{v} \times \mathbf{H}/c).$

We took the magnetic viscosity $\eta_m \equiv c^2/(4\pi\sigma)$, as well as the conductivity $\sigma$ and adiabatic index $\gamma$, to be constant throughout the computational domain, and to not vary with time. The magnetic field in the simulation region had one of three configurations: a poloidal field ($H_z$, $H_r$), a toroidal field $H_\varphi$, or a superposition of the two. With the aim of ensuring that the divergence of the magnetic field was precisely zero, $\nabla \cdot \mathbf{H} = 0$, we used the toroidal component $A_\varphi$ of the magnetic vector potential $\mathbf{A}$ in the computations instead of the poloidal component of the magnetic field, comprised of $H_z$ and $H_r$, $\mathbf{H} = \nabla \times \mathbf{A}$. The system of equations (5)–(8) in cylindrical coordinates in terms of the toroidal component of the magnetic vector potential $A\varphi$ and the toroidal component of the magnetic field $H_\varphi$ has the following form [29]:

$$\frac{\partial(r\rho)}{\partial t} + \frac{\partial(r\rho v_z)}{\partial z} + \frac{\partial(r\rho v_r)}{\partial r} = 0, \quad (10)$$

$$\frac{\partial(r\rho v_z)}{\partial t} + \frac{\partial}{\partial z}\left[r\left(\rho v_z^2 + p + \frac{H_\phi^2}{8\pi}\right)\right] \quad (11)$$

$$+ \frac{\partial(r\rho v_z)}{\partial r} = -\frac{1}{4\pi}\frac{\partial(rA_\phi)}{\partial z}\left(\nabla^2 A_\phi - \frac{A_\phi}{r^2}\right),$$

$$\frac{\partial(r\rho v_r)}{\partial t} + \frac{\partial(r\rho v_z v_r)}{\partial z} \quad (12)$$

$$+ \frac{\partial}{\partial r}\left[r\left(\rho v_r^2 + p + \frac{H_\phi^2}{8\pi}\right)\right] = \rho v_\phi^2 + p$$

$$- \frac{H_\phi^2}{8\pi} - \frac{1}{4\pi}\frac{\partial(rA_\phi)}{\partial r}\left(\nabla^2 A_\phi - \frac{A_\phi}{r^2}\right),$$

$$\frac{\partial M_\phi}{\partial t} + \frac{\partial(M_\phi v_z)}{\partial z} + \frac{1}{r}\frac{\partial(rM_\phi v_z)}{\partial r} \quad (13)$$

$$= \frac{1}{4\pi}\left(\frac{\partial(rA_\phi)}{\partial r}\frac{\partial H_\phi}{\partial z} - \frac{\partial A_\phi}{\partial z}\frac{\partial(rH_\phi)}{\partial r}\right),$$

$$\frac{\partial A_\phi}{\partial t} + v_z\frac{\partial A_\phi}{\partial z} + v_r\frac{1}{r}\frac{\partial(rA_\phi)}{\partial r} \quad (14)$$

$$= \eta_m\left(\nabla^2 A_\phi - \frac{A_\phi}{r^2}\right),$$

$$\frac{\partial H_\phi}{\partial t} + \frac{\partial(v_z H_\phi)}{\partial z} + \frac{\partial(v_r H_\phi)}{\partial r} \quad (15)$$

$$= \frac{\partial}{\partial z}\left(\frac{v_\phi}{r}\frac{\partial(rA_\phi)}{\partial r}\right) - \frac{\partial}{\partial r}\left(v_\phi\frac{\partial A_\phi}{\partial z}\right)$$

$$+ \eta_m\left(\nabla^2 B_\phi - \frac{H_\phi}{r^2}\right),$$

$$\frac{\partial(r\rho\varepsilon)}{\partial} + \frac{\partial(r\rho\varepsilon v_z)}{\partial z} + \frac{\partial(r\rho\varepsilon v_r)}{\partial r} \quad (16)$$

$$= -p\left(\frac{\partial(rv_z)}{\partial z} + \frac{\partial(rv_r)}{\partial r}\right)$$

$$+ \frac{r\eta_m}{4\pi}\left[\left(\frac{1}{r}\frac{\partial(rH_\phi)}{\partial r}\right)^2 + \left(\frac{\partial H_\phi}{\partial z}\right)^2\right.$$

$$\left. + \left(\nabla^2 A_\phi - \frac{A_\phi}{r^2}\right)^2\right].$$

Here, we have used the notation $\eta^m \equiv c^2/(4\pi\sigma)$ =const for the magnetic viscosity and $M_\varphi \equiv \rho v_\varphi r$ for the angular momentum. The poloidal components of the magnetic field can be expressed $H_r = -\partial A_\varphi/\partial z$ and $H_z = (1/r)\partial(rA_\varphi)/\partial r$. We solved the system (10)–(16) using an axially symmetric, hybrid type MHD difference scheme with finite conductivity based on the local-interaction method [30] and the flux-correction method [31, 32]. This scheme was first applied to the solution of astrophysical problems in [33, 34], and was then further developed in [35, 36].

### 4.4. Boundary and Initial Conditions

The system (5)–(8) was solved numerically in the cylindrical region ($0 \leq z \leq Z_{max}$, $0 \leq r \leq R_{max}$). We assumed that there was magnetic field, but no currents, at the outer boundaries. We imposed this using so-called free boundary conditions; i.e., for any quantity $f$, $\partial f/\partial n = 0$. Here, $\partial f/\partial n$ denotes the derivative along the normal to a surface; i.e., at the outer boundary, $f(r, z) =$ const. In accordance with the assumption of axial symmetry, the boundary conditions along the $z$ axis ($r = 0$, $0 \leq z \leq Z_{max}$) can be written

$v_r = v_\varphi = 0$, $A_\varphi = H_\varphi = 0$.

Fixed boundary conditions were specified at the boundary ($z = Z_{min}$, $R_d \leq r \leq R_{max}$), $\rho = \rho_0$ and $v_0 = 0$, and the flow of matter through the boundary into the simulation region was prohibited. "Free" boundary conditions ($\partial/\partial n = 0$) were specified at the opposite boundary ($z = Z_{max}$, $0 \leq r \leq R_{max}$), which enable matter to flow out of the region. The inflow of matter through this boundary into the simulation region was likewise prohibited. Free boundary conditions ($\partial/\partial n = 0$) were also specified at the cylindrical boundary ($Z_{min} \leq z \leq Z_{max}$, $r = R_{max}$). We checked the influence of the boundary conditions on the solutions obtained by carrying out a series of computations with various sizes for the simulation region.

The magnetic vector potential $A = A_\varphi$ was determined and specified throughout the computational domain. For a uniform magnetic field $H$ directed along the $z$ axis, the vector potential can be written $A_\varphi = rH$, $A_r = 0$, $A_z = 0$, and the toroidal component of the field is $H_\varphi = 0$. For a toroidal magnetic field, $A_\varphi = 0$ everywhere, and the toroidal component of the field can be written $H_\varphi = H_0 r/R_d$ for $0 \leq r \leq R_d$ and $H_\varphi = H_0 R_d/r$ for $R_d \leq r \leq R_{max}$, where $H_0$ is a constant.

At the initial time $t = 0$, the magnetic field was purely poloidal or toroidal. The initial density of the matter was taken to be uniform throughout the simulation region, apart from at the target, and to have a low value $\rho = \rho_0$. Apart from at the target, the initial velocity of the matter was zero throughout the simulation region, $v_z = v_r = 0$. The matter had no angular momentum, $v_\varphi = 0$. In the vicinity of the target ($0 \leq R \leq R_d$, $0 \leq z \leq Z_d$), we specified the density to be $\rho \_ \rho_0$, and the velocity of particles ejected from the target was determined from the condition
$\rho v^2/2 = E, v = \sqrt{2E/\rho},$ where $E$ is the energy at the initial time.

### 4.5. Dimensionless Parameters

After reducing the system (5)–(8) to dimensionless form, a set of characteristic dimensionless parameters arises. The first parameter

$$\beta = \frac{8\pi P_{0jet}}{H_0^2} = \frac{2}{\gamma}\frac{c_{s0}^2}{V_{A0}^2}, \qquad (17)$$

determines the relationship between the characteristic values of the matter pressure $P_{0jet} = \rho_{0jet} c_{s0}^2/\gamma$ and the magnetic energy density. Here, $\rho_{0jet}$ is the density of the inflowing matter at the boundary of the computational domain, $P_{0jet}$ and $c_{s0}$ are the pressure and sound speed in the inflowing matter, $H_0$ is the magnetic field in the region, and

$$V_{A0} = \frac{H_0}{\sqrt{4\pi \rho_{0jet}}}$$

is the Alfvén velocity of the inflowing matter at the boundary of the computational domain. The second parameter of the numerical model is the characteristic dimensionless magnetic viscosity (or the characteristic magnetic Reynolds number $Re_M$, which is its inverse):

$$\tilde{\eta}_M = \frac{\eta_M}{L_0 V_{A0}} = \frac{c^2}{4\pi\sigma L_0 V_{A0}} = \frac{1}{Re_M}, \quad (18)$$

where $\sigma$ is the constant effective conductivity of the plasma, $L_0$ is the characteristic size, and $V_{A0}$ the characteristic Alfvén velocity used to determine the magnetic Reynolds number $Re_M$. We adopted the scale $R_{max}$ for the characteristic size $L_0$, and the Alfven velocity of the inflowing matter at the computational boundary $V_{A0}$ for the characteristic velocity $V_0$.

## 4.6. Results of Numerical Simulations

**4.6.1. Hydrodynamical approximation.** A hydrodynamical approximation with $H = 0$ is usually used to test numerical MHD models. This enables estimation of the correctness of the choice of sizes for the simulation region, the parameters, and the boundary conditions. Apart from the vicinity of the target, the initial matter density was uniform and low throughout the simulation region, $\rho = \rho_0$, and the initial velocity of the matter was zero throughout the simulation region, $v_z = v_r = 0$. The matter had no angular momentum, $v_\varphi = 0$.

In the vicinity of the target ($0 \leq R \leq R_d$, $0 \leq z \leq Z_d$), the density was specified to be $\rho \approx 300\rho_0$ and the supersonic velocity of the particles ejected from the target to be $v \geq c_s$, where $c_s = \gamma P_0/\rho_0$ is the sound speed in the unperturbed region. For simplicity of the test computations, the velocity of the ejected matter had only a $z$ component at the initial time, $v_z$, with $v_r = 0$. The initial distributions of the density and velocity are shown in Fig. 3 together with the results of the simulations of the ejection of matter from the target in the hydrodynamical approximation. The color scale corresponds to the logarithm of the matter density, the arrows show local velocity vectors, and the length of an arrow is proportional to the corresponding speed. Under the action of the thermal energy of the laser beam, the target material begins to move along the $Z$ axis toward the opposite wall of the chamber, expanding in the $R$ direction as it propagates. A shock is formed, which moves from the target toward the end of the chamber, leaving a stable, conical-shaped outflow behind it. There is modest collimation of the matter due to the external (background) gas. Figure 4 shows the distribution of the flow density along the $R$ axis at various distances $z$ from the target surface at time $t = 120$. The thin solid curve corresponds to the flow density at the edge of the target, the dotted curve to the density at the center of the chamber, $z = 2.5$, and the dot–dashed curve to the density near the end of the chamber, $z = 5.0$. The flow density at the surface of the target is maximum at the axis, corresponding to the initial conditions. At the end of the chamber, at $z = 5.0$, the density is maximum in a ring extending from $R = 1.6$ to $2.0$. Figure 5 shows the distribution of the kinetic energy along the $R$ axis at various distances $z$ from the target surface at time $t = 120$. The thin solid, dotted, and dot–dashed curves corresponds to the flow density at the edge of the target, at the center of the chamber, $z = 2.5$, and near the end of the chamber, $z = 5.0$, respectively. The kinetic energy of the plasma at the target surface is maximum at the axis, corresponding to the initial conditions. At the end of the chamber, at $z = 5.0$, the kinetic energy is maximum in a ring extending from $R = 1.6$ to $2.0$. We conclude from Figs. 3–5 that there is modest collination of the matter by the external gas, even in the absence of a magnetic field, and weak ring structures in the plasma distribution can be seen at the chamber wall opposite to the target.

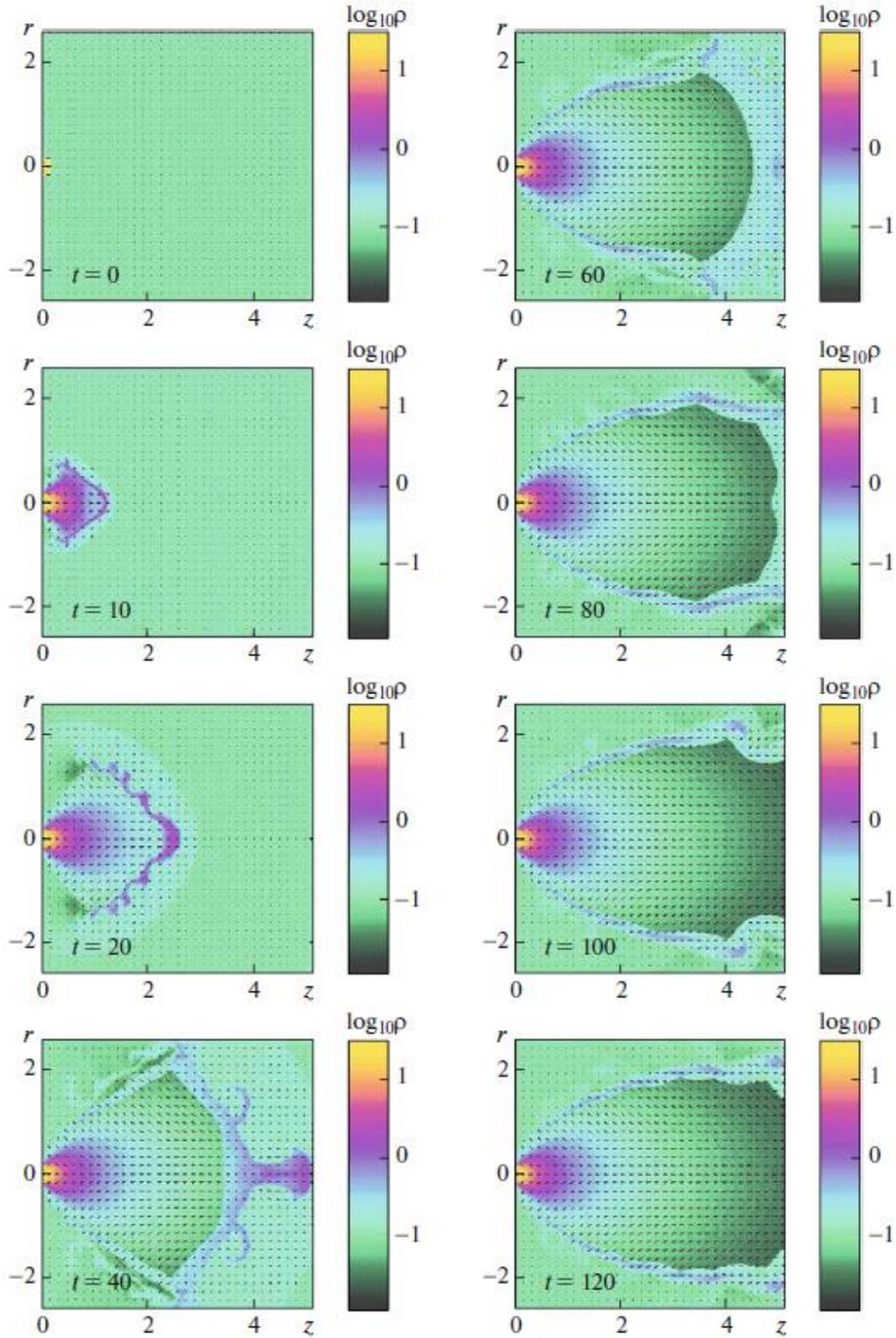

**Fig. 3.** Matter flow pattern for the hydrodynamical case at times $t = 0, 10, 20, 40, 60, 80, 100,$ and $120$ ($t$ given in arbitrary units).

**4.6.2. Poloidal magnetic field $\beta = 10^{-3}$.** In the next stage of our simulations, we included a constant, external poloidal magnetic field $H$ directed along the normal to the initial plane of the target. We conducted a series of computations for various values of the parameter $\beta$. The characteristic parameter values for this case were

$$\beta = \frac{8\pi P_\infty}{H_0^2} = 10^{-3}, \tilde{\eta}_M = \frac{1}{\text{Re}_M} = 10^{-5}$$

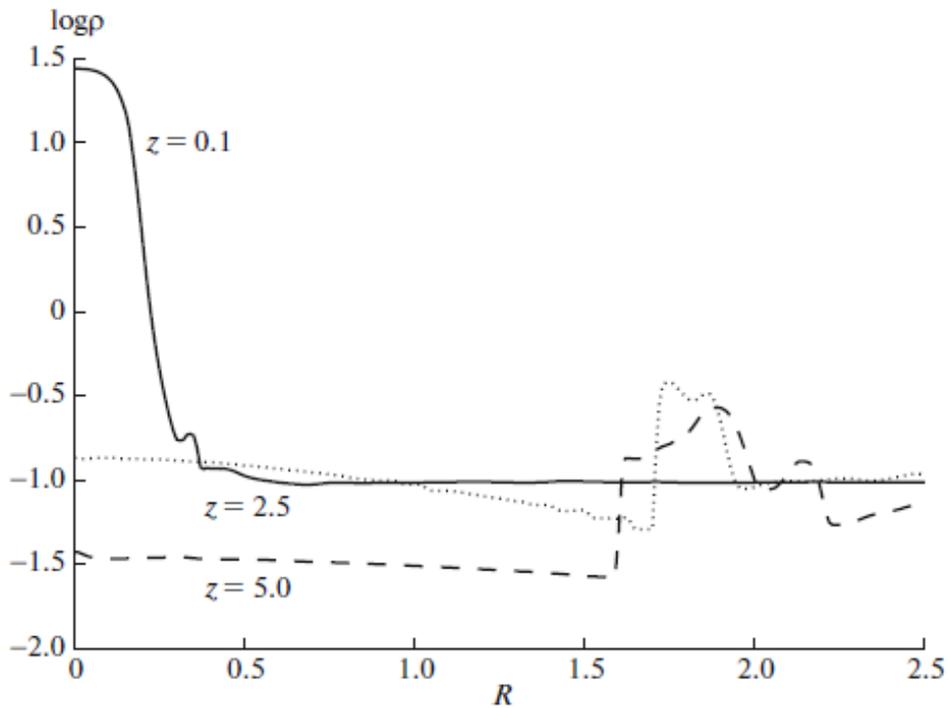

**Fig. 4.** Flow density along $R$ as a function of distance from the target in the hydrodynamical approximation at time $t = 120$ (see text).

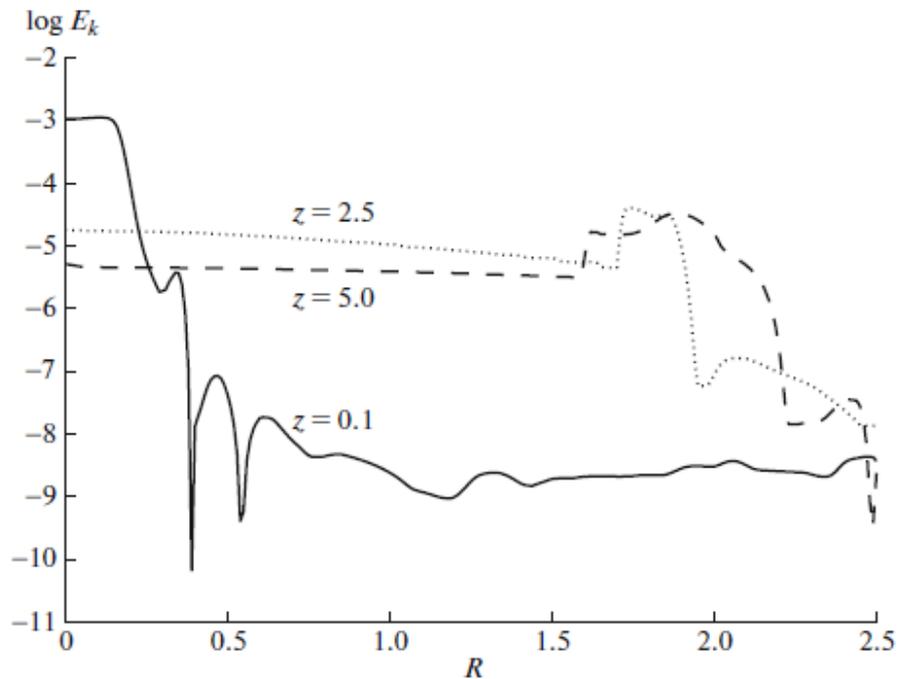

**Fig. 5.** Distribution of kinetic energy along $R$ as a function of the distance to the target in the hydrodynamical approximation at time $t = 120$ (see text).

Apart from the vicinity of the target, the matter density was initially uniform and low throughout the simulation region, $\rho = \rho_0$, and the initial velocity of the matter was zero throughout the simulation region, $v_z = v_r = 0$. The matter had no angular momentum, $v_\varphi = 0$. In the vicinity of the target ($0 \leq R \leq R_d$, $0 \leq z \leq Z_d$), the density was specified to be $\rho \approx 300\rho_0$ and the supersonic velocity of the particles ejected from the target to be $v \geq c_s$. For simplicity of the test computations, the velocity of the ejected matter had only a z component at the initial time, $v_z$, with $v_r = 0$. The uniform magnetic field was specified by the relations $A_\varphi = rH$, $A_r = 0$, $A_z = 0$, $H_\varphi = 0$.

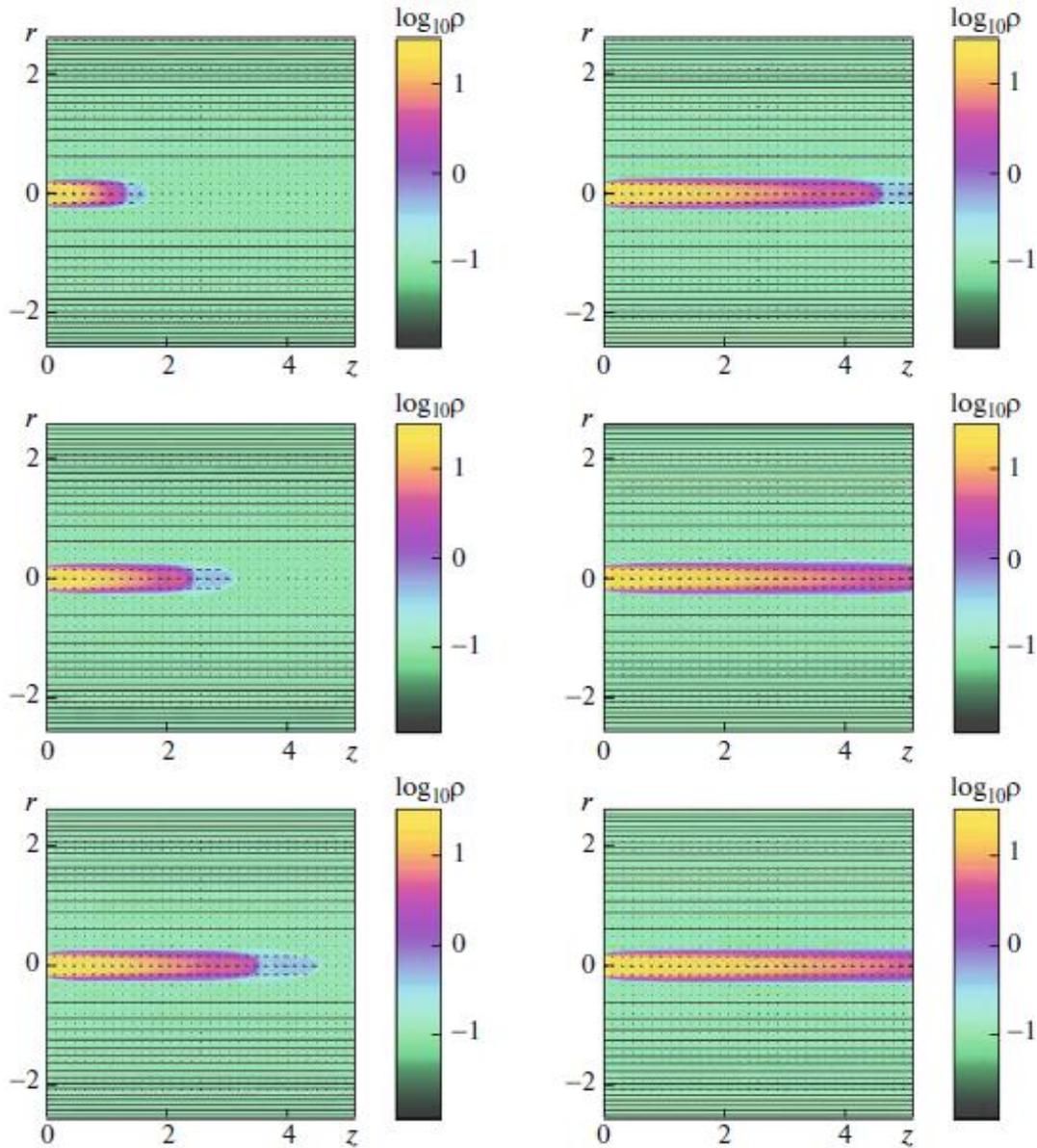

**Fig. 6.** Flow pattern of the matter in the MHD case $\beta = 10^{-3}$ at times $t = 5, 10, 15$ (left), 20, 25, and 30 (right).

The results of the simulations of the ejection of material for the case $\beta = 10^{-3}$ are shown in Fig. 6. The color scale corresponds to the logarithm of the matter density. The thin black curves show the magnetic-field lines and the arrows the local velocity vectors; the lengths of the arrows are proportional to the corresponding speeds. Under the action of the thermal energy of the laser beam, the ejected target material begins to move along the Z axis, toward the end of the chamber. An elongated shock forms and moves from the target toward the end of the chamber, leaving behind it a stable outflow. The flow is well collimated due to the presence of the strong external magnetic field, and the matter does not expand appreciably in the $R$ direction. Collimation by the external (background) gas is not important in the presence of the strong poloidal magnetic field.

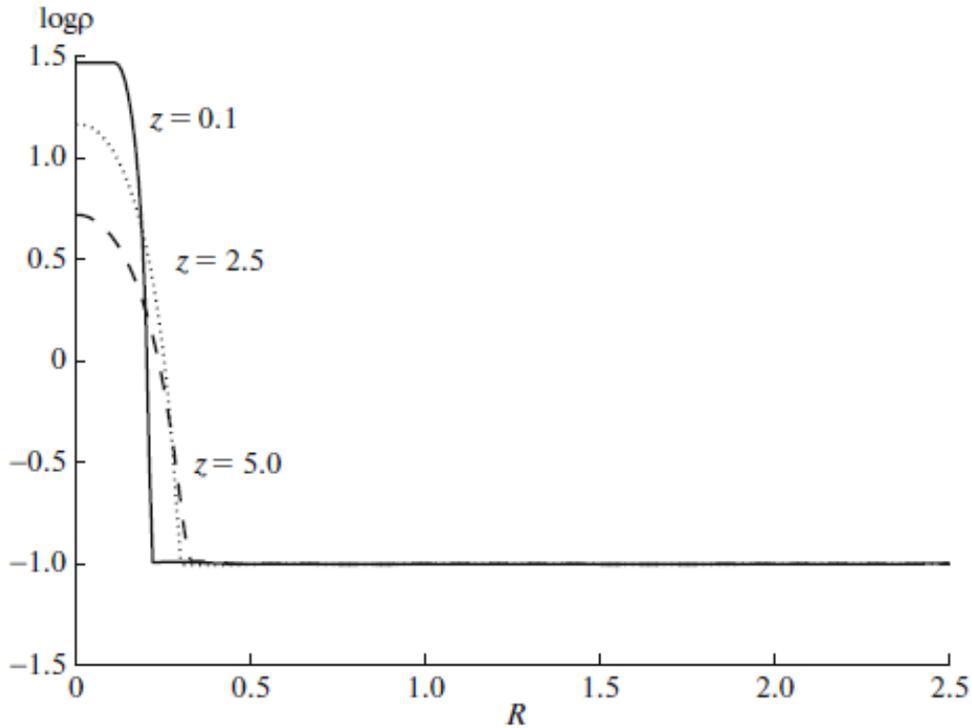

**Fig. 7.** Flow density in the $R$ direction as a function of distance to the target for the case $\beta = 10^{-3}$ at time $t = 30$ (see text).

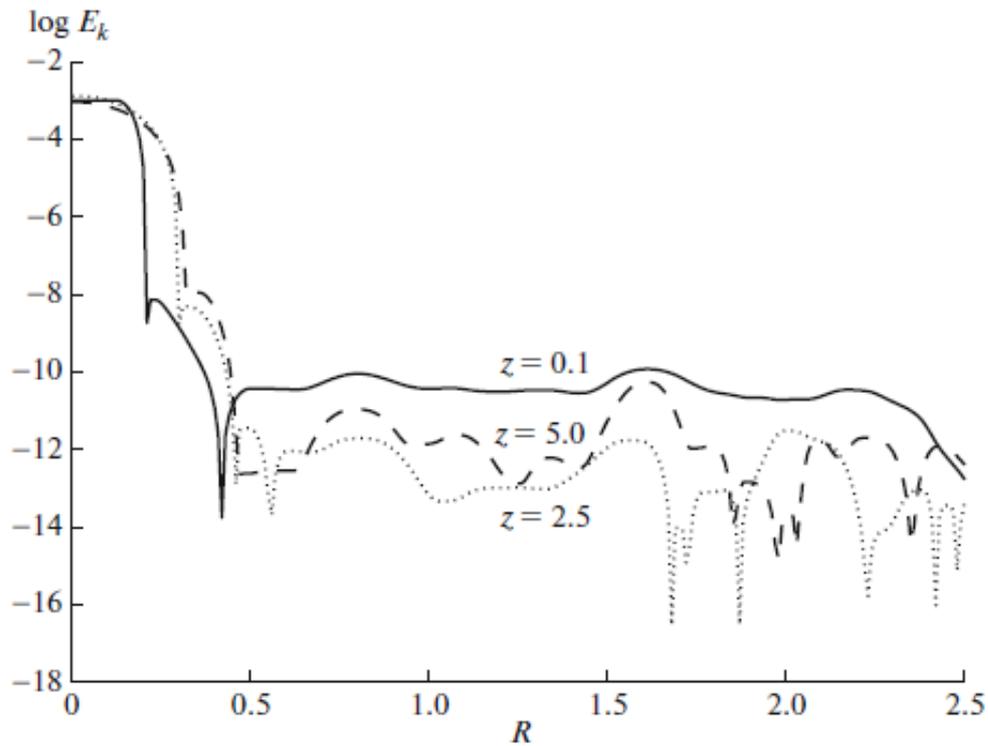

**Fig. 8.** Distribution of the kinetic energy in the $R$ direction as a function of distance to the target for the case $\beta = 10^{-3}$ at time $t = 30$ (see text).

Figure 7 shows the distribution of the flow density along the $R$ axis at various distances $z$ from the target surface at time $t = 30$. The thin solid, dotted, and dot-dashed curves correspond to the flow density at the edge of the target, at the center of the chamber ($z = 2.5$), and near the end of the chamber ($z = 5.0$), respectively. At any distance from the target, the maximum plasma density is located in a narrow range of $R$ corresponding to the size of the target. Figure 8 shows the distribution of the kinetic energy along the $R$ axis at various distances $z$ from the target surface at time $t = 30$. The thin solid, dotted, and dot-dashed curves correspond to the flow density at the edge of the target, at the center of the chamber ($z = 2.5$), and near the end of the chamber ($z = 5.0$), respectively. At any distance from the target, the kinetic energy is maximum within the ejection. We conclude from Figs. 6–8 that, in the presence of a fairly strong poloidal magnetic field, a fairly well defined continuous spot in the plasma distribution is observed at the chamber wall opposite to the target.

**4.6.3. Poloidal magnetic field $\beta = 10^{-1}$.** In this computation, we decreased the magnetic-field strength. The characteristic parameters for this case were

$$\beta = \frac{8\pi P_\infty}{H_0^2} = 10^{-1}, \tilde{\eta}_M = \frac{1}{\mathrm{Re}_M} = 10^{-5}.$$

Apart from the vicinity of the target, the matter density was initially uniform and low throughout the simulation region, $\rho = \rho_0$, and the initial velocity of the matter was zero throughout the simulation region, $v_z = v_r = 0$. The matter had no angular momentum, $v_\varphi = 0$.

In the vicinity of the target ($0 \leq R \leq R_d$, $0 \leq z \leq Z_d$), the density was specified to be $\rho \approx 300\rho_0$ and the supersonic velocity of the particles ejected from the target to be $v \geq cs$. For simplicity of the test computations, the velocity of the ejected matter had only a $z$ component at the initial time, $v_z$, with $v_r = 0$. The uniform magnetic field was specified by the relations $A_\varphi = rH$, $A_r = 0$, $A_z = 0$, $H_\varphi = 0$. The initial time corresponds to the case $\beta = 10^{-3}$. The results of the simulations of the ejection of material for the case $\beta = 10^{-1}$ are shown in Fig. 9. The color scale for the left column corresponds to the logarithm of the matter density, and the color scale for the right column to the logarithm of the temperature. The thin black curves show the magnetic-field lines. After it is ejected, the matter begins to expand in all directions, but the flow is then collimated by the magnetic field. The influence of the external gas on the collimation of the flow is insignificant in this case. A shock forms in the flow at a distance $z \approx 1.6$. Figures 10 and 11 show the dependence of the Mach number $M = v/cs$ on the distance from the target $z$ and the distance from the axis $R$. Both figures correspond to time $t = 45$. The velocity of the matter grows to the distance $z \approx 1.6$, reaching the maximum Mach number $M \approx 10$. After the passage of the shock, the matter velocity falls sharply to $M \approx 0.5$, then again begins to grow to $M \approx 2$ at $z \approx Z\max$.

Figure 12 shows the distribution of the flow density in the $R$ direction for various distances $z$ from the target at time $t = 45$. The solid, dot-dashed, dashed, and dashed curves show the flow density at the edge of the target, at $z = 1.2$, at the center of the chamber ($z = 2.5$), and near the end of the chamber ($z = 5.0$), respectively. The density distribution in the plasma flow depends on the distance to the target. While the maximum plasma density in the flow initially occupies a small range of $R$ values comparable to the size of the target, the density distribution forms a well defined ring structure in the vicinity of the shock, $z = 1.2$.

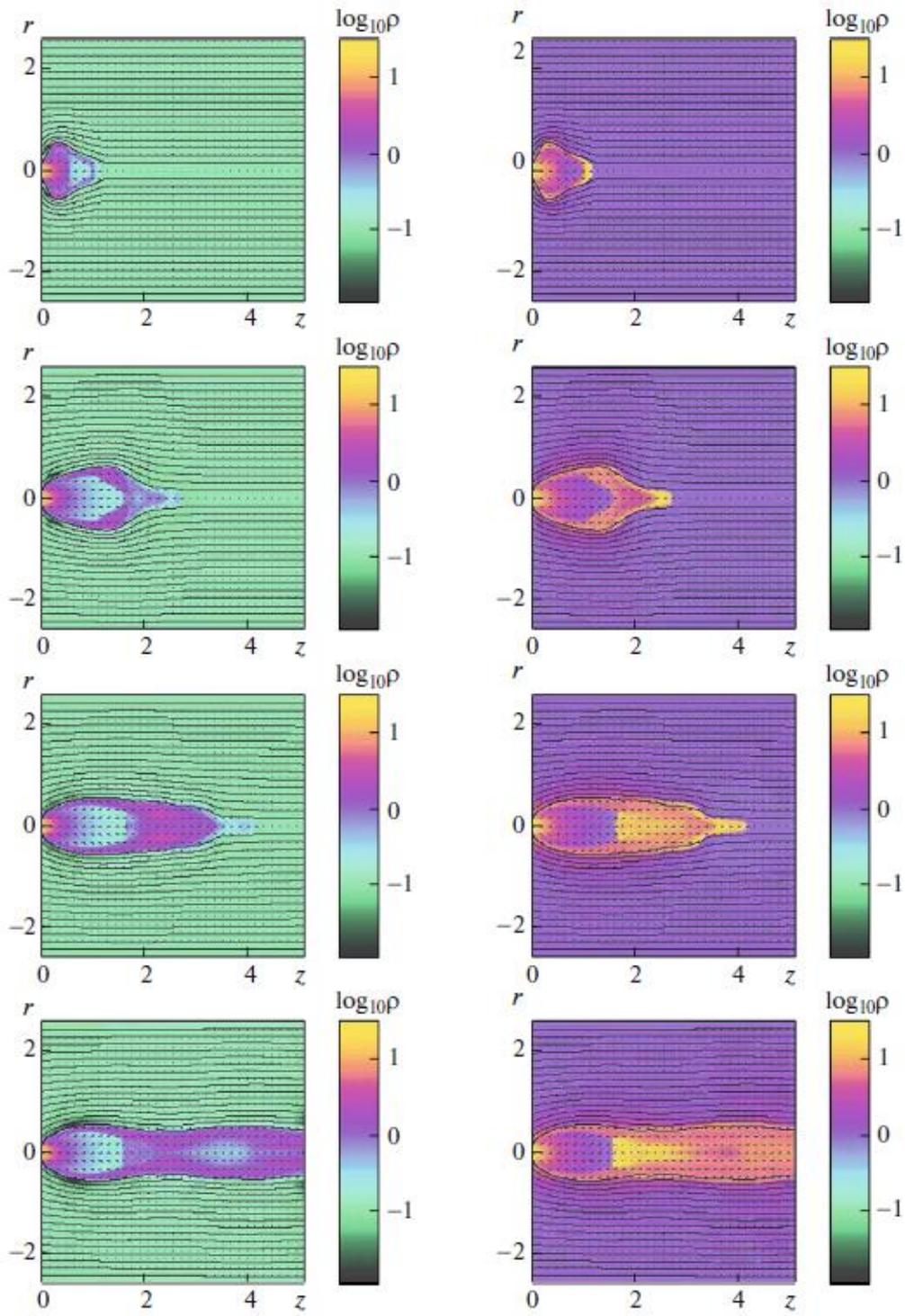

**Fig. 9.** Matter flow pattern in the case of a poloidal magnetic field with $\beta = 10^{-1}$ at times $t = 5$, 10, 25, and 45 (from top to bottom).

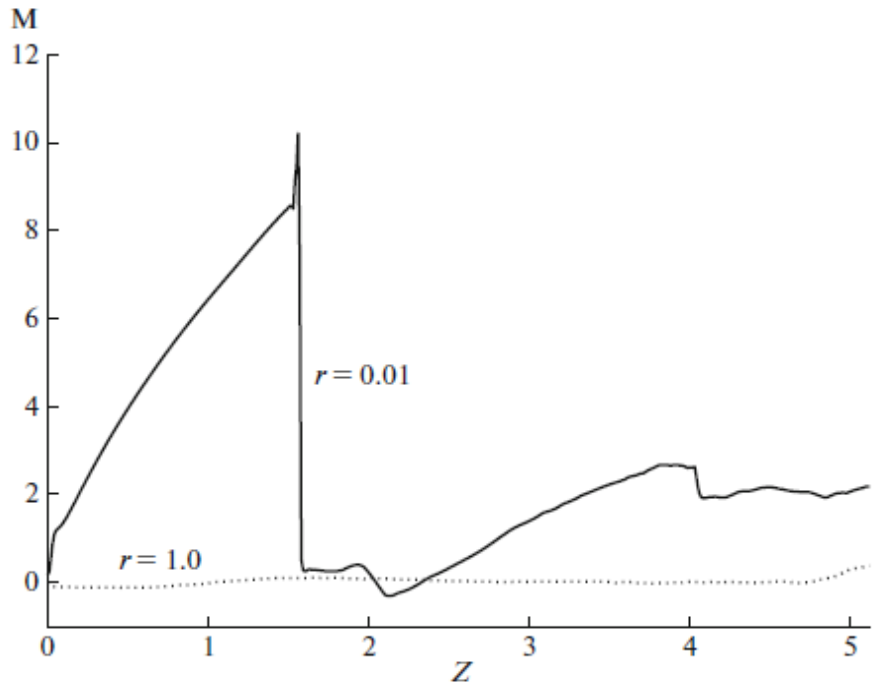

**Fig. 10.** Mach number $M = v/c_s$ as a function of the distance $z$ from the target for $\beta = 10^{-1}$ at time $t = 45$.

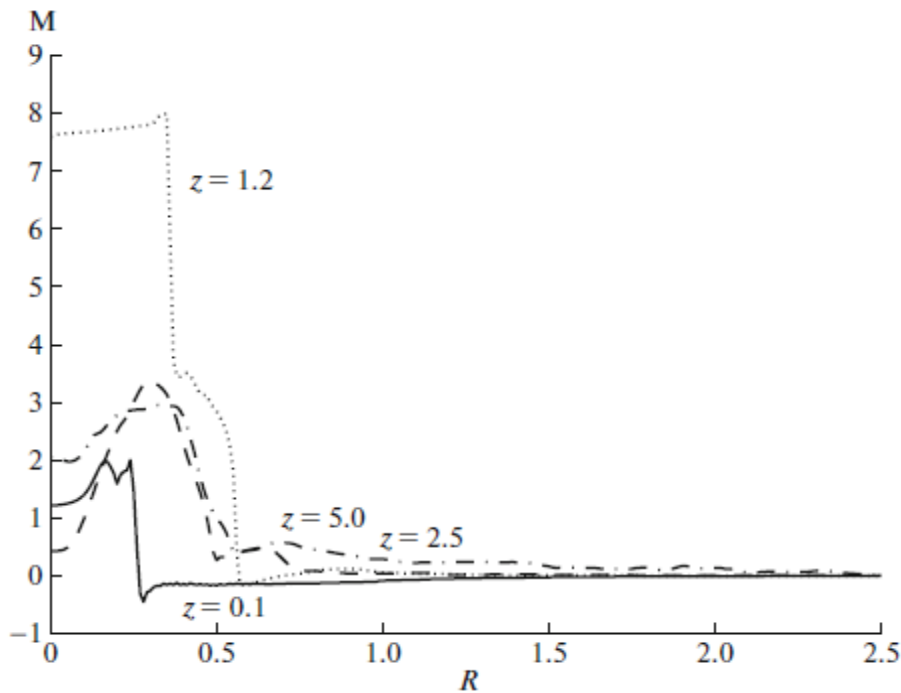

**Fig. 11.** Mach number $M = v/c_s$ as a function of the distance to the axis $R$ for $\beta = 10^{-1}$ at time $t = 45$.

After the passage of the shock, the density distribution becomes more uniform, and the ring structure becomes blurred. Figures 10 and 12 can be used to determine the opening angle of the conical flow at the distance of the shock, which is approximately $\theta \approx 40°$.

Figure 9 shows the formation of a cavity delineating the shock near the target, whose size depends on the strength of the poloidal magnetic field. Similar results were obtained in [5]. This leads to the formation of ring structures, which are observed in the experiment (see below).

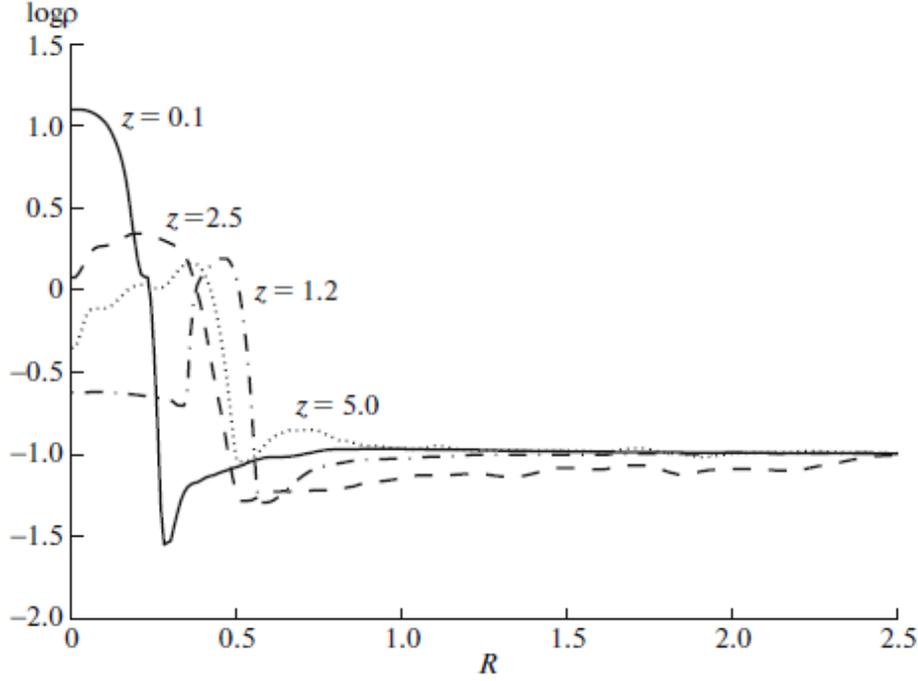

**Fig. 12.** Matter flow density along $R$ as a function of the distance from the target $z$ for the case $\beta = 10^{-1}$ at time $t = 45$.

The opening angle of these structures depends on the strength of the magnetic field. This angle is large when the magnetic field is weak (and also in the absence of a field, see Fig. 3). When the field is increased, the transverse size of the cavity, and accordingly the opening angle of the ring structure, decreases, and further a continuous spot is observed in place of a ring (Fig. 6).

### 4.6.4. Toroidal magnetic field $\beta = 10^{-1}$. In

In this computation, we changed the configuration of the magnetic field. The characteristic parameters for this case were

$$\beta = \frac{8\pi P_\infty}{H_0^2} = 10^{-1}, \quad \tilde{\eta}_M = \frac{1}{\text{Re}_M} = 10^{-5}.$$

Apart from the vicinity of the target, the matter density was initially uniform and low throughout the simulation region, $\rho = \rho_0$, and the initial velocity of the matter was zero throughout the simulation region, $v_z = v_r = 0$. The matter had no angular momentum, $v_\varphi = 0$.

In the vicinity of the target ($0 \leq R \leq R_d$, $0 \leq z \leq Z_d$), the density was specified to be $\rho \approx 300\rho_0$ and the supersonic velocity of the particles ejected from the target to be $v \geq c_s$. For simplicity of the test computations, the velocity of the ejected matter had only a $z$ component at the initial time, $v_z$, with $v_r = 0$. In a simple model, the magnetic field can be defined as the field of a conductor with radius $Rd$ carrying a current directed along the Z axis: $A_\varphi = 0$ everywhere, $H_\varphi = H_0 r/R_d$ for $0 \leq r \leq R_d$, and $H_\varphi = H_0 R_d/r$ for $R_d \leq r \leq R_{\max}$, where $H_0$ is a constant. The initial time corresponds to the case with poloidal field $\beta = 10^{-1}$. These simulation results are shown in Fig. 13. The color scales in the left and right columns correspond to the logarithm of the matter density and the logarithm of the temperature.

Figure 13 shows an absence of collimation of the flow and of the ring structures in this case. This may be associated with insufficiencies of our chosen method for specifying the toroidal field as the field of a conductor carrying a current in the $Z$ direction. In this case, the magnetic field has its maximum value $H_0$ at the edge of the target $R_d$, then falls off with $R$, $H_\varphi = H_0 R_d/r$. Such a magnetic field is not strong enough to collimate the plasma flow, and the collimation effect of the external gas is also disrupted. As a result, the plasma expands more isotropically even than in the case with no magnetic field. It is obvious that the magnetic field in the experiment does not have either a toroidal or a poloidal configuration, but instead a much more complex superposition of fields.

## 5. EXPERIMENTAL PART

Experiments aimed at studying the spatial distributions of beams of accelerated protons using CR-39 track detectors were carried out at the Neodim 10- TW picosecond laser facility [37]. This laser installation has the following parameters for the laser pulses produced: energy up to 10 J, wavelength 1.055 $\mu$m, duration 1.5 pc, and contrast of the laser radiation of order $10^7$. The focusing system based on an off-axis parabolic mirror with a focal length of 20 cm provides concentration of no less than 40% of the laser-beam energy into a spot with a diameter of 15 $\mu$m and a corresponding peak intensity of $2 \times 10^{18}$ W/cm2.

A schematic of the setup for the experiments designed to investigate the spatial distributions of beams of accelerated protons is shown in Fig. 14. The CR-39 track detectors with sizes of 25×25 mm$^2$ were located 20 mm from the target, along the normal in front of the target (A), along the normal behind the target (B), and along the direction of the laser radiation behind the target (C). Filters of Al with thicknesses of 11–80 $\mu$m were installed in front of the CR-39 track detectors. The target was a foil of Cu with a thickness of 30 or 50 $\mu$m or a foil of Ta with a thickness of 50 $\mu$m. The results of the experiments showed that the proton beams were registered only in the normal direction, in front of the target (A) and behind the target (B), with the proton beam being more collimated in the latter case. We present the results of experiments on the spatial distribution of proton beams accelerated outward from the rear surface of the target.

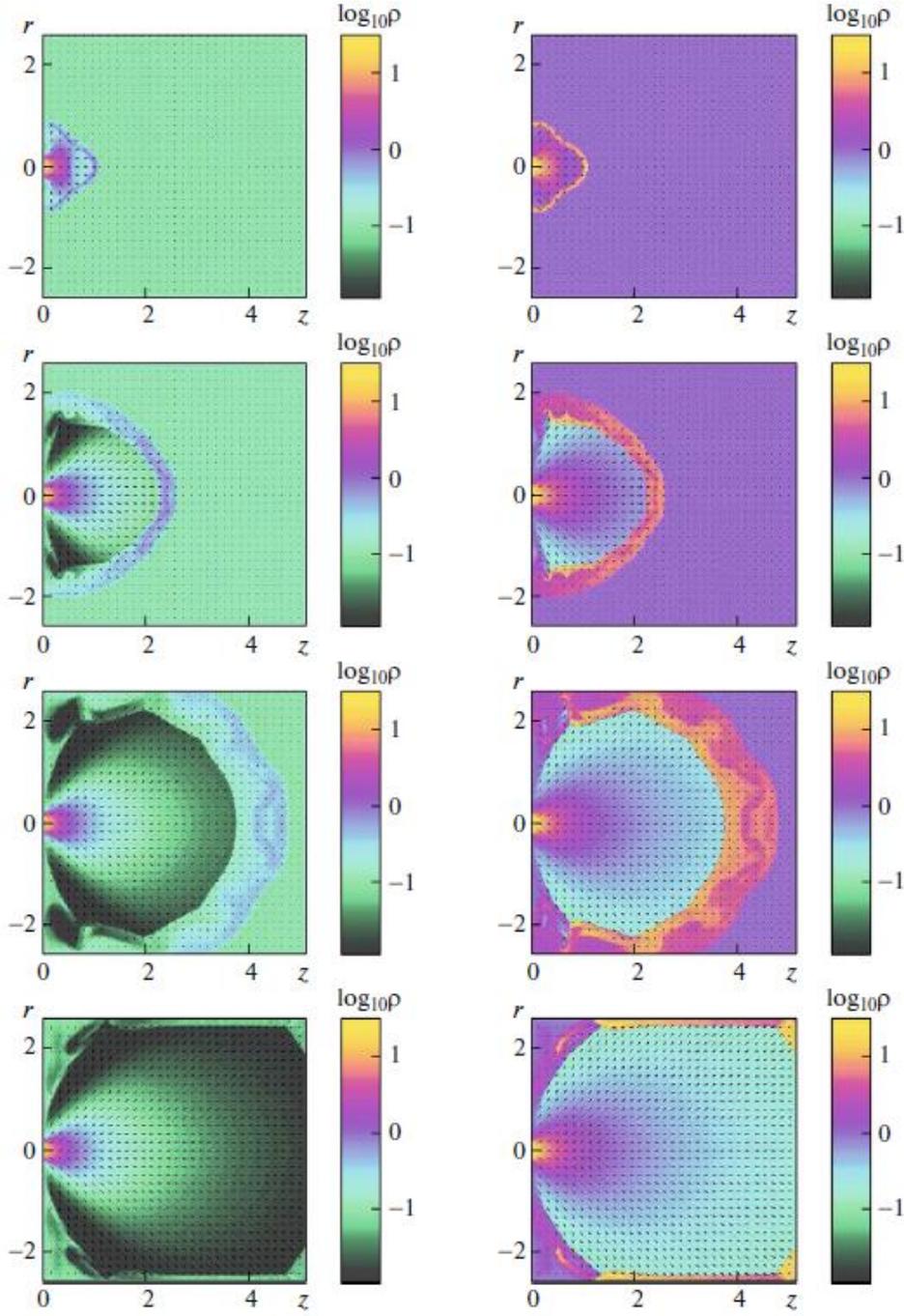

**Fig. 13.** Matter flow pattern in the case of a toroidal magnetic field for the case $\beta = 10^{-1}$ at times $t = 5, 10, 25,$ and $45$ (from top to bottom).

Figure 15 shows images of proton beams with various energies obtained at the CR-39 track detectors when a Cu foil target with thickness 50 μm was used as the target. When the proton-beam energy is increased from 0.8 to 1.7 MeV, the divergence of the proton beam decreases from 21.8° to 15.4°.

Figure 16 shows images of proton beams with the same energy ($E_p > 1.7$ MeV) at the CR-39 track detectors when a Cu foil with thicknesses of 30 and 50 μm and a Ta foil with thickness 50 μm were used as targets. When the thickness of the Cu foil target is increased from 30 to 50 μm, the divergence for proton beams with the same energy decreases from 20.5° to 15.4°. In the transition from one type of target (Cu, 50 μm) to another with the same thickness but a higher

atomic number (Ta, 50 μm; $Z_{Cu} = 29$, $Z_{Ta} = 73$), the proton-beam divergence decreases from 15.4°
to 5.7°.

Figure 17 presents the experimental dependences of the proton-beam divergences on the proton energy for targets of Cu (30 μm and 50 μm) and Ta (50 μm), and for energies from 0.8 to 3 MeV.

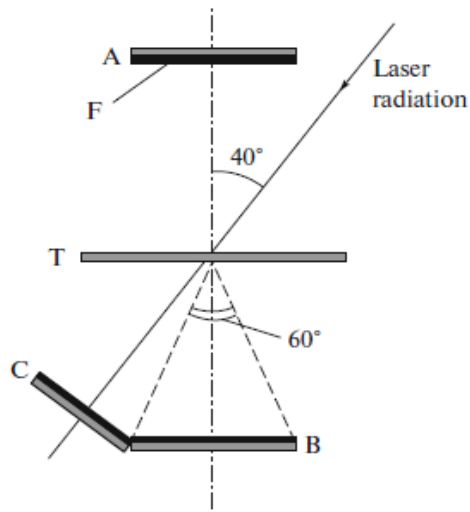

**Fig. 14.** Scheme for the experiments investigating the spatial distribution of beams of accelerated protons. A, B, C are the CR-39 track detectors, F filters of Al with thicknesses of 11–80 μm, and T the target of Cu (30 and 50 μm) or Ta (50 μm).

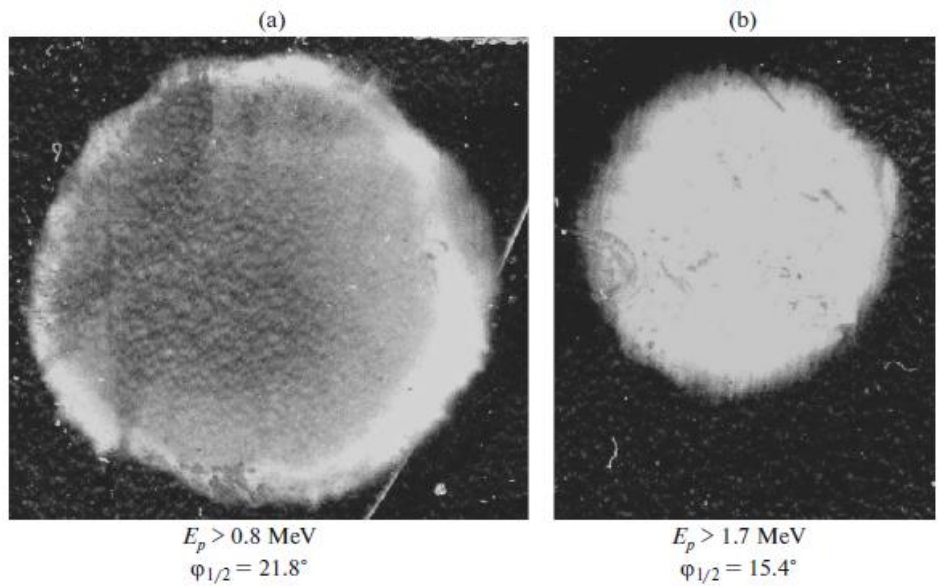

**Fig. 15.** Images of proton beams with (a) energies $E_p > 0.8$ MeV and $\phi_{1/2} = 21.8°$, and with (b) $E_p > 1.7$ MeV and $\phi_{1/2} = 15.4$

These data can be used to draw conclusions about the angular distribution of beams of protons accelerated outward from the rear surface of the target. The proton-beam divergence decreases for protons with higher energy and for targets with higher thickness and higher atomic number. We also studied this angular distribution using a lower laser intensity, $10^{18}$ W/cm². In this case, the divergence for proton beams of a given energy decreases with the laser intensity. For example, Fig. 18 shows images of proton beams with the same energy ($E_p > 0.8$ MeV) at the CR-39 track detectors for a Cu target with thickness 30 μm obtained for laser intensities of $2 \times 10^{18}$ W/cm² (Fig. 18a) and $10^{18}$ W/cm² (Fig. 18b). This decrease in the laser intensity by a factor of two leads to a decrease in the proton-beam divergence by nearly a factor of two, from 26.4° to 14°. The maximum proton energy was reduced from 5 to 2.5 MeV in this case.

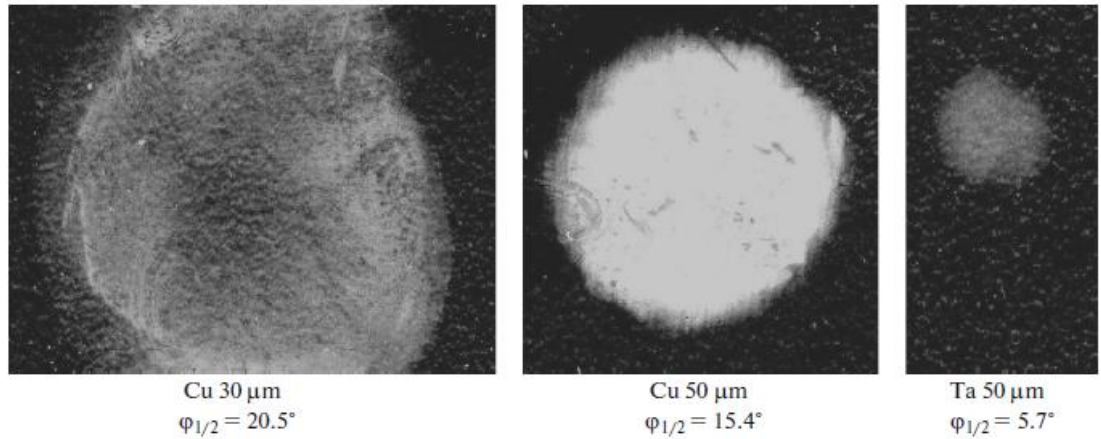

**Fig. 16.** Images of proton beams with the same energy ($E_p > 1.7$ MeV) at the CR-39 track detectors for various targets (from left to right): Cu with thickness 30 μm ($\phi_{1/2} = 20.5°$), Cu with thickness 50 μm ($\phi_{1/2} = 15.4°$), and Ta with thickness 50 μm ($\phi_{1/2} = 5.7°$).

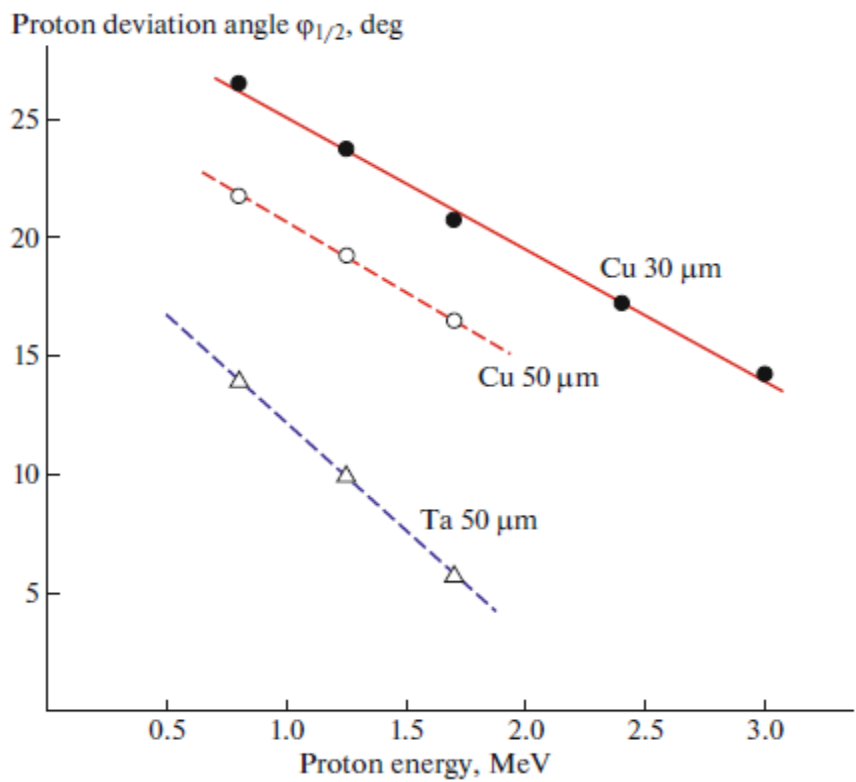

**Fig. 17.** Dependence of the proton-beam divergence on the proton energy for targets of Cu (thicknesses of 30 and 50 μm) and Ta (thickness of 50 μm).

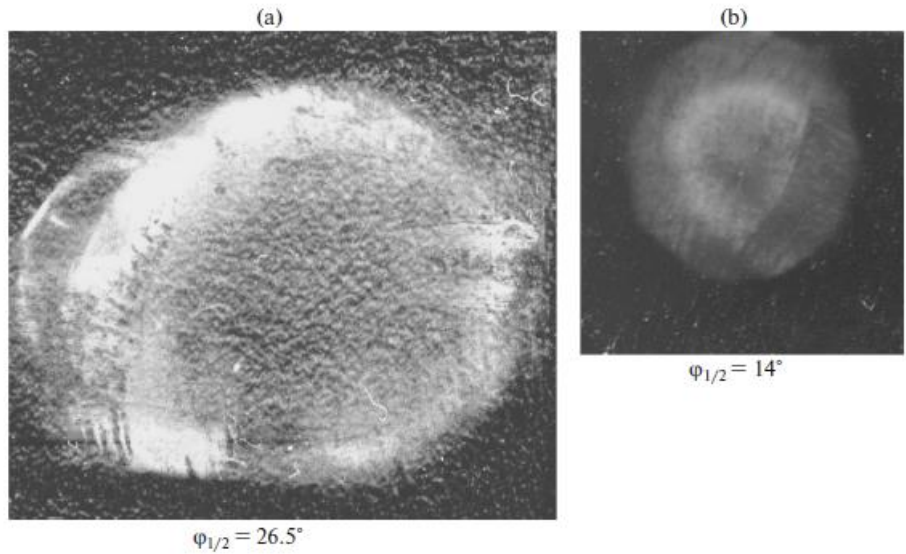

**Fig. 18.** Images of proton beams with the same energy ($E_p > 0.8$ MeV) at the CR-39 track detectors for a Cu target (thickness 30 μm), obtained for laser intensities of (a) $I = 2 \times 10^{18}$ W/cm$^2$ ($\phi_{1/2} = 26.5°$) and (b) $10^{18}$ W/cm2 ($\phi_{1/2} = 14°$).

We also note an interesting fact about the beam distributions for protons with relatively low energies. As can be seen in Figs. 15a, 18a, and 18b, ring structures formed by tracks of protons with large diameters (low energies) are clearly visible in the images for beams with proton energies exceeding 0.8 MeV.

Figure 19 presents distributions for protons with energies exceeding 0.8 MeV inside this ring structure, with a Cu target with thickness 30 μm. Figure 19a shows the distribution for all protons with energies of 0.8–3 MeV and Fig. 19b the distribution for protons with energies of 0.8–1.7 MeV. The size of the image where the track density is maximum is about 10 mm, and the proton-beam divergence is $\phi_{1/2} \approx 14°$. The distribution of tracks in the image is not uniform: tracks with small diameters (made by protons with high energies) dominate at the center, where the proton flux is maximum. Then, the well defined boundary of a ring formed by tracks with larger diameters (lower-energy protons) is visible about 3 mm from the center ($\phi_{1/2} \cong 8.5°$).

Experimental studies aimed at investigating the angular distributions of fast protons generated in interactions of powerful laser radiation with thin targets were carried out earlier in [38].

Ring structures in proton fluxes accelerated from the rear surface of a target were observed earlier in [39–42]. Ring structures in proton fluxes propagating along the normal to the target from both the front and rear surfaces of the target were observed in [39]. When a slab of Al with thickness 125 μm was used as a target with a laser intensity of $5 \times 10^{19}$ W/cm2, protons with a maximum energy of about 30 MeV were registered. It follows from these results that the propagation of protons with relatively modest energies of about 3 MeV occurs in a cone with an opening angle of about +30°. When the proton energy is increased, the diameters of the ring structures decrease, and the protons propagate in a narrower cone. The number of fast protons with energies exceeding 2 MeV was $10^{12}$ particles per laser pulse. Krushelnick et al. [39] suggested that the origin of the ring structures in the proton fluxes emerging from the rear surface of the target is an azimuthal magnetic field of ≈30 MG inside the target, which is generated by fluxes of fast electrons in the target. Ring structures in proton fluxes from the rear surface of plane targets with thicknesses of 5–100 μm subject to a laser intensity of $5 \times 10^{18}$ W/cm$^2$ were observed in [40]. The experimental results indicate that the diameter of these ring structures decreases as the proton energy and target thickness increase. The output of protons with energies exceeding 1 MeV was about $2 \times 10^9$ per laser pulse. Murakami et al. [40] explain the presence of ring structures in the proton fluxes as due to the influence of a quasi-stationary toroidal magnetic field with a strength of several tens of MG, which is generated by a flux of fast electrons emerging from the rear of the target.

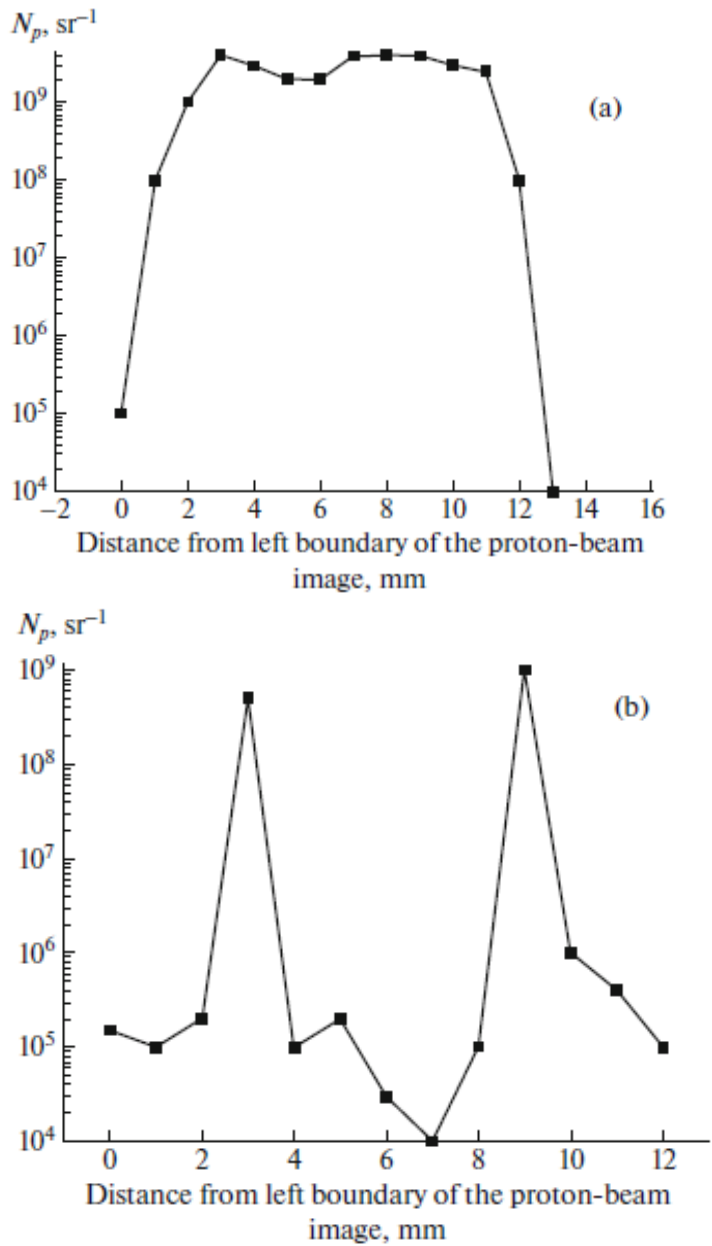

**Fig. 19.** Distribution of the proton flux within the ring structure, obtained for a Cu target (thickness 30 μm), for protons with energies of (a) 0.8–2.5 MeV and (b) 0.8– 1.7MeV.

Ring structures in proton fluxes from the rear surface of an Al target with thickness 100 μm subject to a laser intensity of $8 \times 10^{19}$ W/cm$^2$ were observed in [41]. Experiments with Al targets that were both cool and heated to 300–400°Cwith deposits of PEEK polyethyline coatings on both their front and rear surfaces were used to identify the proton-acceleration mechanism. Well defined ring structures for protons with energies exceeding 6 MeV and divergences of +30° were observed for both the cool and heated targets. When the heated targets were used, protonring structures were observed only when when the PEEK polyethylene coatingwas deposited only on the front surface of the target. Based on the results of such experiments, Zepf et al. [41] concluded that the protons forming the ring structures are generated at the front surface of the target, and that the reason for the appearance of these structures is a magnetic field generated inside the target by a beam of electrons propagating in the target. The results of measurements of the output of MeV protons and deuterons when thin targets are irradiated by picosecond laser pulses with a mean laser intensity of ≤ 4 × 1018 W/cm2 are presented in [42]. A ring structure formed by traveling ions and a very narrow (0.5°) angular divergence for the ion beam were observed in experiments with a Be target with thickness 12 $\mu$m. Andreev et al. [42] suggest that the appearance of the ring structure in the proton fluxes is due to inhomogeneity in the electron charge of the transverse component of the ambipolar field and the magnetic field at the rear surface of the target. Theoretical estimates of the energy and flight angles for ions are close to the experimental values, testifying that the ions were generated at the rear side of the foil.

We will now compare the results of our experiments on the spatial distribution of beams of accelerated protons with the results of [39–42].

First, in both our experiments and the experiments of [39–42], ring structures in the proton fluxes generated from the rear surface of the target were observed.

Second, in both our experiments and the experiments of [39–42], the divergence of the proton beam is +30° for proton energies of several MeV. The divergence of the proton beam decreases as the proton energy and the target thickness increase, and as the laser intensity decreases. We have also shown that the divergence of the proton beam decreases with increasing atomic number of the target material.

Third, only in [42] was a proton ring structure with a very narrow divergence of order +0.5° detected. Note that there is currently no generally accepted theory for the formation of the observed ring structures in the proton fluxes generated when powerful laser radiation interacts with the target material. Therefore, the detailed studies of the distribution of beams of protons accelerated by laser intensities of $10^{18}$–$10^{20}$ W/cm$^2$ using targets with various thicknesses and atomic numbers presented in this section will undoubtedly provide useful input for theories of the formation of spatial structures in the distribution of the accelerated protons.

## 6. CONCLUSION

We have presented the results of numerical simulations of magnetized and unmagnetized supersonic jets. A substantial expansion of the jet arises in the absence of a magnetic field. In the presence of a strong poloidal magnetic field, the jet expands only weakly, confirming the possibility of magnetic collimation of astrophysical jets. The toroidal magnetic fields we have considered proved to be insufficiently strong to collimate the jets.

Our computations show that the appearance of ring structures is possible for some values of the poloidal magnetic field. We have also presented the results of laboratory laser experiments at the Neodim installation. Clearly distinguishable ring structures formed by protons with energies of 0.8–1.7 MeV are observed in the experimental results. The opening angle for these protons (≈40°) corresponds roughly to the results of the numerical simulations.

Thus, our comparison of the results of our laboratory experiment, our numerical simulations of magnetized jets, and the results of [5] confirm the possible formation of ring structures whose characteristics depend on the magnetic-field strength. Further theoretical and experimental studies in this area are required. The formation and development of jets in laboratory laser experiments is a complex physical situation including a large number of different physical processes. The creation of numerical simulations of such experiments taking into account all possible physical processes is a complex mathematical task, and remains difficult to realize. We have applied an MHD approximation as a first step in the studies we have considered here.

## ACKNOWLEDGMENTS

O.D. Toropina thanks V.V. Savel'ev (Institute of Applied Mathematics, Russian Academy of Sciences) for presentation of the original version of the MHD simulation program. This work was partially supported by the Russian Foundation for Basic Research (grants 14-29-06045, 16-02-00350, 17-02- 00021, 17-02-00760) and a grant of the President of the Russian Federation in support of Leading Scientific Schools (NSh-6579.2016.2) and program I.7P.